\documentclass[a4paper]{aa}

\usepackage{graphicx}
\usepackage{amsfonts}
\usepackage{amsmath}
\usepackage{longtable}

\newcommand{\msun}{M_{\odot}}
\newcommand{\oversim}[2]{\protect{\mbox{\lower0.5ex\vbox{%
   \baselineskip=0pt\lineskip=0.2ex
   \ialign{$\mathsurround=0pt #1\hfil##\hfil$\crcr#2\crcr\sim\crcr}}}}}
\newcommand{\simgreat}{\mbox{$\,\mathrel{\mathpalette\oversim>}\,$}} % >~ sign
\newcommand{\simless} {\mbox{$\,\mathrel{\mathpalette\oversim<}\,$}} % <~ sign

\begin{document}

\title{The Pleiades mass function~: models
  versus observations}

\author{E. Moraux, P. Kroupa \& J. Bouvier}

\author{E. Moraux\inst{1,2}\and P. Kroupa\inst{3,4,5}\and J. Bouvier\inst{1}}

\offprints{E. Moraux}

\mail{moraux@ast.cam.ac.uk}

\institute{Laboratoire d'Astrophysique, Observatoire de Grenoble, B.P.
53, 38041 Grenoble Cedex 9, France
\and Institut of Astronomy, Cambridge, CB3 0HA, UK
\and Institut f\"ur Theoretische Physik und Astrophysik der Universit\"at
Kiel, D-24098 Kiel, Germany
\and Sternwarte der Universit\"at Bonn, Auf dem H\"ugel 71, D-53121
Bonn, Germany
\and {\it Heisenberg Fellow}}

\date{Accepted 24/06/2004}

\authorrunning{E. Moraux et al.}  
\titlerunning{The Pleiades mass function: models versus observations}

%%%%%%%%%%%%%%%%%%%%%%%%%%%%%%%%%%%%%%%%%%%%%%%%%%%%%%%%%%%%%%%%%%%%%

\abstract{Two stellar-dynamical models of binary-rich embedded
  proto-Orion-Nebula-type clusters that evolve to Pleiades-like
  clusters are studied with an emphasis on comparing the stellar mass
  function with observational constraints. By the age of the Pleiades
  (about 100~Myr) both models show a similar degree of mass
  segregation which also agrees with observational constraints. This
  thus indicates that the Pleiades is well relaxed and that it is
  suffering from severe amnesia. It is found that the initial mass
  function (IMF) must have been indistinguishable from the standard or
  Galactic-field IMF for stars with mass $m\simless 2\,M_\odot$,
  provided the Pleiades precursor had a central density of about
  $10^{4.8}$~stars/pc$^3$. A denser model with $10^{5.8}$~stars/pc$^3$
  also leads to reasonable agreement with observational constraints,
  but owing to the shorter relaxation time of the embedded cluster it
  evolves through energy equipartition to a mass-segregated condition
  just prior to residual-gas expulsion. This model consequently looses
  preferentially low-mass stars and brown dwarfs (BDs), but the effect
  is not very pronounced. The empirical data indicate that the
  Pleiades IMF may have been steeper than Salpeter for stars with
  $m\simgreat 2\,M_\odot$.
  }

\maketitle

%%%%%%%%%%%%%%%%%%%%%%%%%%%%%%%%%%%%%%%%%%%%%%%%%%%%%%%%%%%%%%%%%%%%%

\section{Introduction}

Because of its richness and proximity ($\sim120$~pc, Robichon et
al. 1999), the Pleiades cluster with an age of $120$ Myr (Stauffer et
al. 1998; Mart\'{\i}n et al. 1998) is one of the best studied young
open clusters. It has a mass of about $740\,M_\odot$, contains 
$\sim$1000 stars and has a tidal radius of about 13.1~pc, a half mass
radius of about 3.66~pc (Pinfield et al. 1998), and a binary-rich
population with properties similar to that in the Galactic field
(Bouvier et al. 1997). Recent stellar-dynamical computations suggest
that the Pleiades-precursor may have been very similar to the about
1~Myr old Orion Nebula cluster (ONC) (Kroupa, Aarseth \& Hurley 2001,
hereinafter KAH), opening the possibility of seeing the same kind of
object at two very different evolutionary stages.

Several surveys have been performed to date to measure its stellar
mass function which is now well constrained from the stellar to the
sub-stellar domain. Recent estimates extend as deep as $0.03\msun$
(Moraux et al. 2003). A pressing issue in star formation is to know
whether this mass function (MF) observed at an age of $\sim$100 Myr is
representative of the initial mass function (IMF). In other words, is
the observed Pleiades population representative of the cluster
population at the time it formed and how did it evolve within 100
Myr~?

KAH present calculations of the formation of Galactic clusters using
Aarseth's {\sc Nbody6} variant {\sc GasEx} (Aarseth 1999). They
defined an initial model from a set of assumptions describing the
outcome of the star formation process and studied how it could lead to
the formation of the ONC and then to the Pleiades by dynamical and
stellar evolution. They found encouraging consistency with
observational constraints on the radial density profile, kinematics
and binary properties for both the ONC and the Pleiades. With this
contribution we compare the MF predicted by the models with currently
available data to test if the Pleiades IMF may have been compatible
with the ``standard'' IMF (Kroupa 2001) or with the log-normal IMF
found to match the Galactic field population down to the substellar
regime (Chabrier 2002).

In this paper, we describe briefly KAH's models in section~2 before
comparing them to current estimates of the Pleiades MF and presenting
relevant results about the radial distribution and the evolution of
the cluster population in section~3. The conclusions are given in
section~4.

\section{Model description and predictions}

KAH develop stellar-dynamical models of cluster formation under
realistic conditions. Their computations are performed with Aarseth's
{\sc Nbody6}-variant {\sc GasEx} which allows accurate treatment of
close encounters without softening and multiple stellar systems in
clusters, incorporates stellar evolution and a Galactic tidal field in
the solar neighbourhood. In the models the initial stellar and
sub-stellar masses are distributed following the standard, or
Galactic-average, three-part power-law IMF $\displaystyle \xi(m)
\propto m^{-\alpha_{\rm i}}$, where $dn=\xi(m)\,dm=\xi_{\rm
L}(m)\,d{\rm log}_{10}m$ is the number of objects in the mass interval
$m,m+dm$ and log$_{10}m$, log$_{10}m+d$log$_{10}m$, respectively.
$\xi_{\rm L}(m)=(m\,{\rm ln}10)\,\xi(m)$ is the ``logarithmic
IMF''. The standard IMF has (Kroupa 2001)
\begin{equation*}
 \alpha_{0}= +0.3, \,\,\,\, 0.01\le m< 0.08\msun ;%$$
\end{equation*}
\begin{equation}
 \alpha_{1}= +1.3, \,\,\,\, 0.08\le m< 0.50\msun ;%$$
  \label{eq:stmf}
\end{equation}
\begin{equation*}
 \alpha_{2}= +2.3, \,\,\,\, 0.50\le m\le 50\msun.\,\,\,\,%$$
\end{equation*}
Initially all objects, stars and brown dwarfs, are assumed to be in
binary systems with component masses chosen randomly from the IMF. The
total binary proportion
$$ f_{\rm tot} = \frac{N_{\rm bin}}{N_{\rm bin}+N_{\rm sing}}, $$
where $N_{\rm bin}$ and $N_{\rm sing}$ are the number of binary and
single-star systems respectively, is then equal to unity. The models
initially consist of $10^4$~stars and brown dwarfs in 5000 binaries.

Two models are constructed, both with a spherical Plummer density
profile for the stars and the gas initially. The half-mass radius is
$R_{0.5}=0.45$~pc for model~A with an initial central number density
$\rho_{\rm C}=10^{4.8}$~stars/pc$^3$, and $R_{0.5}=0.21$~pc for
model~B with $\rho_{\rm C}=10^{5.8}$~stars/pc$^3$. Both models are
embedded in a gas potential with twice the mass in stars and with the
same density profile as the stellar component.  The gas is removed due
to the action of the O stars after 0.6~Myr and the dynamical evolution
is integrated for 150~Myr.  By assuming a star-formation efficiency of
33~per cent and removing the residual gas within a thermal time-scale,
the initially embedded cluster (the proto-ONC) expands rapidly,
matches the ONC by about 1~Myr, and by 100~Myr it forms a bound star
cluster which resembles the observed Pleiades and which forms the
nucleus of an expanding association.

The MF measured in the Pleiades today may bear witness of these
events, because if the embedded cluster was significantly mass
segregated prior to gas removal, then mostly the low-mass stars will
have been lost, and the Pleiades would then be deficient in
very-low-mass stars.

The evolution of the two cluster models is discussed in KAH and
compared to observational constraints on the structure, kinematics and
binary-star properties for both the ONC and the Pleiades cluster. The
core radii as well as the number of stars within these radii predicted
at 0.9 Myr and 100 Myr by both models are in reasonable agreement with
observed values for the ONC and the Pleiades respectively. Available
velocity dispersion measurements are also consistent with the
models. The observed ONC radial density profile is well reproduced by
model B but not model A, whereas for the Pleiades the best fit is
given by model A. Concerning the period distribution for late-type
binaries, both models are consistent with the Pleiades data, although
an intermediate model in term of central density would be better. For
the ONC, only model B leads to agreement with the measured proportion
of binaries with periods $P\ge10^{5}$ days, model A retaining too many
binaries. Overall though, both models fit the ONC and the Pleiades
data quite well, suggesting that the Pleiades cluster most probably
formed from an ONC-like object. Therefore the initial state of the
Pleiades will have been much more concentrated than presently observed
with a half-mass radius between 0.2~and~0.5~pc.

In this paper we extend this study by a comparison to the Pleiades MF
which is now constrained down to the sub-stellar regime. The system
MF which is constructed by counting only the system masses is compared
with the observed MF, because the observational surveys cannot resolve
binary stars. It is interesting to investigate the Pleiades MF to see
if it may have changed and whether it differed from the standard IMF,
given that the search for IMF variations constitutes one of the most
important issues of star formation work.

\section{The observed Pleiades mass function}

\begin{figure}[htbp]
  \centering\leavevmode
  \includegraphics[width=0.9\hsize]{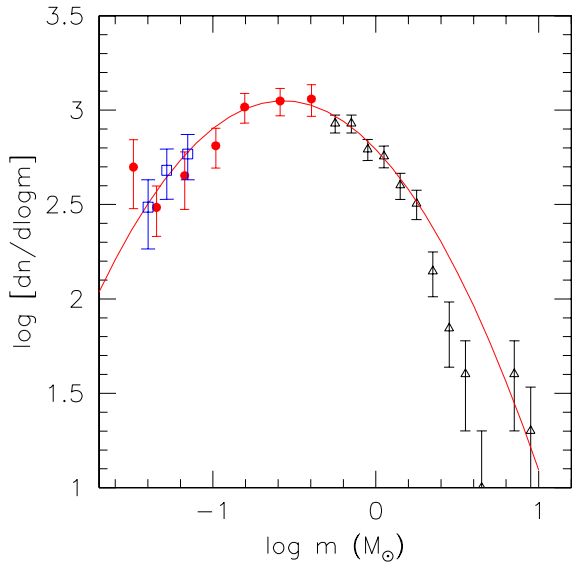}
  \caption{The Pleiades mass function. The open triangles are from the
    Prosser and Stauffer Open Cluster Database and the filled circles
    are the data points from Moraux et al. (2003). The open squares
    have been obtained by applying the empirical magnitude-mass
    relation given by Dobbie et al. (2002). The log-normal fit
    (eq.~\ref{eq:plmf}) found by Moraux et al. (2003) is shown as a
    solid line.}
  \label{mfM78}
\end{figure}

Moraux et al. (2003) performed a deep and extended survey of the
Pleiades cluster using the CFH-12K camera. The survey covered 6.4
sq.deg. reaching up to 3 degrees from the cluster's centre and the
covered mass range extends from 0.03 to 0.45$\msun$. The authors
used the NextGen (Baraffe et al. 1998) and Dusty (Chabrier et
al. 2000) models to convert their luminosity function into a mass
function and they used the Stauffer \& Prosser open cluster
database\footnote{Available at
http://cfa-www.harvard.edu/stauffer/opencl/} to complete their data
for $m\ge 0.4\msun$. They thus computed the mass function from
$0.030\msun$ to $10\msun$ and found that it is reasonably well-fitted
by a log-normal function in logarithmic units (Moraux et al. 2003,
Fig.9)
  \begin{equation}
    \xi_{\rm L}(m) = \frac{dn}{d\log m} \propto \exp\left[ -
      \frac{(\log m - \log m_{0})^2}{2\sigma^{2}}\right]
    \label{eq:plmf}
  \end{equation}
with $m_{0}\simeq0.25\msun$ and $\sigma=0.52$. It is shown on
Fig.~\ref{mfM78}.

However, the lowest mass point of the MF looks a bit different and is
located much above the log-normal fit. Similar features are observed
for other clusters around the same spectral type (M7-M8) and Dobbie et
al. (2002) argued this reflects a sharp local drop in the
luminosity-mass (L-M) relationship, due to the onset of dust formation
in the atmosphere around $T_{eff}=2500$~K. A change in the slope of
the L-M relationship implies that the masses of objects with spectral
types later than M7 may be significantly underestimated by the current
NextGen and Dusty models and that the number of objects in this lowest
mass bin may be overestimated. By applying the empirical
magnitude-mass relation given by Dobbie et al. (2002; their Fig.3) for
the Pleiades low mass stars and brown dwarfs, we find that the lowest
mass point of the Pleiades MF goes down whereas higher mass points go
up (see Fig.~\ref{mfM78}). The log-normal fit does not change however
and becomes even better.

Even if this dust effect has still to be investigated, it suggests
that the lowest mass point of the Pleiades MF may come down in the
future and we consider the log-normal distribution as a good
approximation of the observed MF in this mass domain. 

For $m\simgreat 2\msun$ however, this fit does not seem to be steep
enough. The 2-4$\msun$ MF points are more than 2-sigma below the
log-normal fit.

\section{Comparison models/observations}

\subsection{The mass function}

\begin{figure}[htbp]
  \centering\leavevmode
  \includegraphics[width=0.9\hsize]{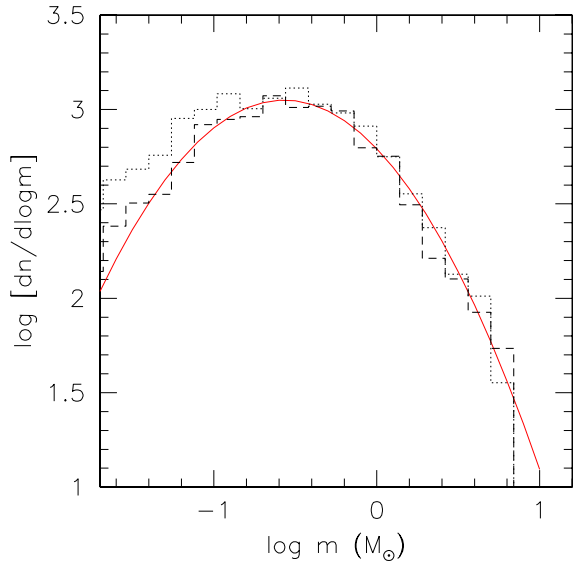}
  \caption{Comparison of the log-normal fit of the observed Pleiades
    MF (shown as a solid line) to the model MFs. The dashed (dotted)
    histogram represents the system mass function at an age of 100~Myr
    from model A (B) of KAH for all systems within R$\le$15 pc of the
    model cluster center. They have been scaled to match the Pleiades
    data around $m\simeq 1\msun$.}
  \label{mfAB}
\end{figure}

The system MFs at $t=100$ Myr resulting from the dynamical and stellar
evolution of KAH's models, assuming random pairing and a binary
fraction $f_{\rm tot}=1$ at $t=0$ are compared to the log-normal fit
of the observed Pleiades MF in Fig.~\ref{mfAB} over the entire mass
range. The system MF from model A (resp. model B) is shown as a dashed
(resp. dotted) histogram.

Both models reproduce well the log-normal distribution for $m\simgreat
0.1\msun$ which indicates that the observed MF (data points in
Fig.~\ref{mfM78}) is also steeper than the model MF for $m\simgreat
2\,M_\odot$. This discrepancy cannot be the effect of the
stellar evolution since the models include its treatment but may
indicate that the true stellar IMF in the Pleiades may have been
steeper than Salpeter ($\alpha=2.35$) above $2\,M_\odot$.

It is useful to note that {\it the KAH models} do not constitute a fit
but {\it are a prediction of the Pleiades MF} in the stellar and the
sub-stellar regime under the assumption that the Pleiades IMF was
identical to the standard IMF (eq.~\ref{eq:stmf}; Kroupa 2001). Our
result here suggests that this may indeed have been the case except
for the possibly steeper IMF for $m\simgreat 2\,M_\odot$.

For $m\simless 0.1\msun$ down to the sub-stellar regime, model A
yields a better fit than model B which suggests that the Pleiades may
not have formed from an embedded cluster as concentrated as model
B. The larger number of isolated BD systems predicted by model B
compared to model A in Fig.~\ref{mfAB} results from the early release
of brown dwarf companions from their primary in the initially denser
environment of model B. The observed period-distribution of late-type
Pleiades binaries seems also to be better reproduced with model~A for
$P\le10^{6}$ days (see KAH, their Fig.11), although an intermediate
model in terms of central density would lead to improved agreement for
longer periods. Therefore, our results tend to favour model A but
model B cannot be excluded since it is still consistent with the
observational result within the uncertainties (especially for the
lower mass bins where the effect of dust formation still needs to be
investigated).

By 100~Myr the binary fraction is not equal to one any longer and
depends on mass. During the early stage of the cluster's life, many
binaries are disrupted because of gravitational interactions. The
binding energy being weak for low mass binaries, many brown dwarfs
(BDs) have been freed and the BD-BD binary fraction predicted by the
models is $\sim20\%$ (see KAH, their fig.9) which is consistent with
current observational results ranging from 15-50\%. Mart\'{\i}n
et al. (2003) find $\sim 15\%\pm^{15}_{5}$ for resolved Pleiades
systems with separations from 7-12 AU but Pinfield et al. (2003)
suggests that the Pleiades total (i.e. unresolved in general) binary
fraction is from 30-50\%. Overall, the observational constraints on
late-type binaries are in reasonable agreement with the models, as
shown by KAH. The binary proportion of the massive stars decreases
first through disruption and then increases as new companions are
captured.

\begin{figure}[htbp]
  \includegraphics[width=0.9\hsize]{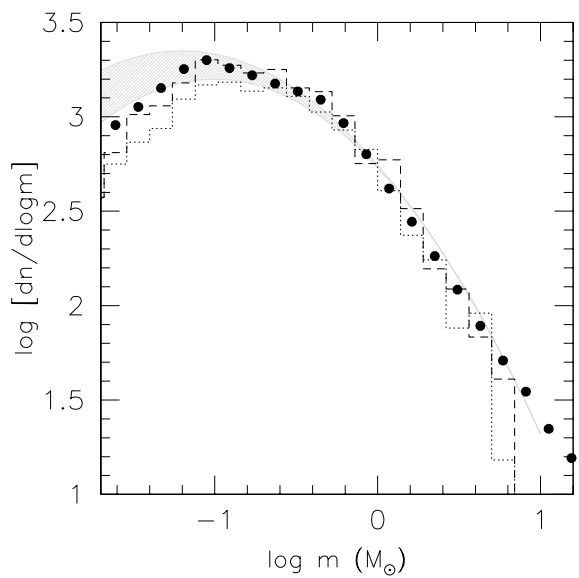}
  \caption{The single object mass function for model A (dashed
  histogram) and model B (dotted histogram) at $t=100$ Myr and
  R$\le$15 pc. The shaded region represents the field mass function
  from Chabrier (2003) with uncertainties. The dots correspond to the
  shape of the standard three-part power-law IMF. All models are
  scaled to agree between $0.5\msun$ and $1\msun$.}
  \label{mf_single}
\end{figure}

The mass function of all the objects counted individually is
represented by a histogram in Fig.~\ref{mf_single} for both models.
It corresponds to the Pleiades MF we would observe if we were able to
resolve all the binaries. If we compare this single star MF with the
standard IMF indicated by the dots, we find that the overall shapes
are similar, although model B shows a marginal deficit of BDs compared
to model A. This indicates that the main difference between the
Pleiades MF observed at an age of $\sim100$ Myr and the Pleiades IMF
is mainly due to the binarity effect and that the cluster dynamical
evolution had little effect on its mass distribution.

The Galactic disk log-normal MF obtained by Chabrier (2003, eq.~17) by
fitting L and T dwarf samples (Chabrier 2002) is found to be slighty
above the histograms in the sub-stellar domain. This suggests that
either the Pleiades IMF is different from the Chabrier's field IMF in
this mass range or that the number of brown dwarfs in the field has
been overestimated. The ratio
\begin{equation*}
R = \frac{N_{obj}(0.02-0.08 \msun)}{N_{obj}(0.15-1 \msun)}
\end{equation*}
where $N_{obj}$ is the number of objects in the respective mass range
is found to be equal to $1.07\pm0.17$ for Chabrier's Galactic disk
log-normal MF, whereas it is equal to $0.81\pm0.05$ for the three-part
power-law standard Pleiades IMF.

\subsection{Mass segregation}

\begin{figure}[htbp]
  \centering\leavevmode
  \includegraphics[width=\hsize]{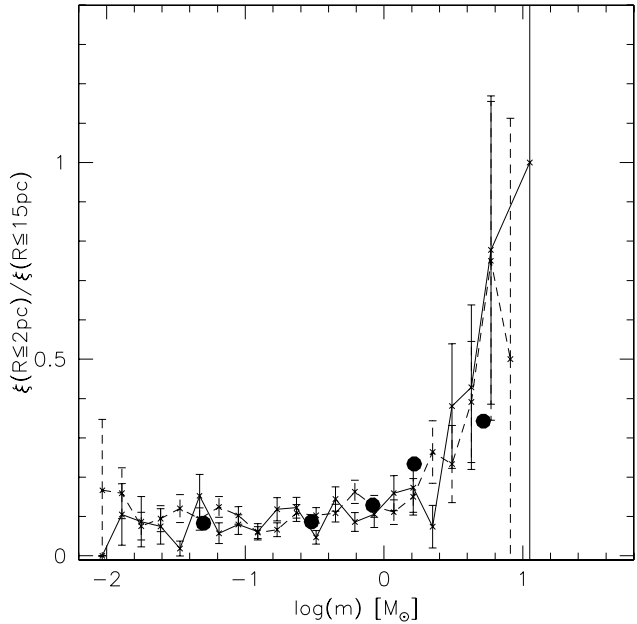}
  \caption{Ratio of the number of systems within R$\le$2 pc to the
  total number of cluster systems (R$\le$15 pc) per mass bin at
  $t=100$ Myr. The solid line corresponds to model A and the dashed
  line to model B. The stellar data points are from the observational
  survey of Pinfield et al. (1998) and the sub-stellar one is from
  Moraux et al.(2003) (filled circles).}
  \label{segreg100}
\end{figure}

In Fig.~\ref{segreg100}, the fraction of systems per mass bin
contained in the cluster centre (R$\le$2 pc) and predicted by the
models at $t=100$ Myr is compared with observational constraints. The
solid line corresponds to model A and the dashed line to the initially
more concentrated model B. The observational results for the Pleiades
are indicated by the filled circles. The data points come from
Pinfield et al. (1998) for the stellar domain and from Moraux et
al. (2003) for the sub-stellar regime. The King density profile found
by these authors for each mass bin has been de-projected, thus
providing the spatial density which has then been integrated between 0
and 2 pc to obtain the number of systems in the inner part of the
cluster.

Model predictions are consistent with observations. The more massive
stars are concentrated around the cluster centre whereas the low mass
objects are much more spread out throughout the cluster.

That the mass segregation at $t=100$ Myr is similar for the two models
indicates the cluster to be completely relaxed at this age, and that
the memory of the initial concentration has been lost. However, if we
compare the models after 1 Myr, mass segregation has not occurred yet
for model A whereas it is already present for model B (see KAH,
fig.~15). This model evolved more rapidly than model~A by virtue of
its larger initial density and thus shorter relaxation time.

\subsection{Evolution of the mass function}

\begin{figure}[htbp]
  \centering\leavevmode
  \includegraphics[width=\hsize]{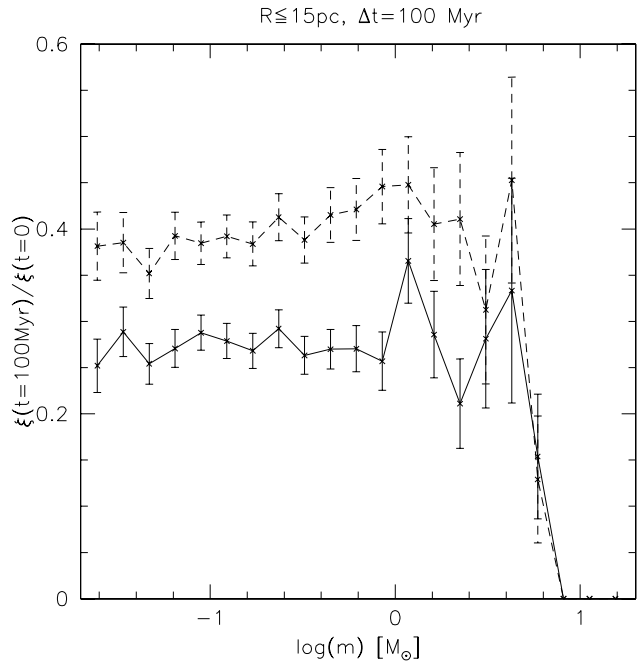}
  \caption{Fraction of individual objects remaining inside the cluster
  (R$\le$15 pc) at $t=100$~Myr compared to the initial number of objects at
  $t=0$. The solid line corresponds to model A and the dashed line to
  model B. The ratio of the individual object MFs is shown, not of the
  system MFs.}
  \label{all_res}
\end{figure}

In the previous sections it has been shown that the Pleiades cluster
is dynamically evolved and that mass segregation has occurred, as is
also found to be the case by Portegies Zwart et al. (2001). Here we
consider the detailed structure of the MF to infer if the preferential
loss of stars of a certain mass may be evident in the models. The
ratio of the present-day MF at $t=100$~Myr to the IMF is plotted for
models~A and~B in Fig.~\ref{all_res}. The comparison between the
cluster population at 100 Myr and the initial cluster population shows
that the evolved population is depleted above about $4\,M_\odot$ due
to stellar evolution. However there are no significant
evolutionary effects that deplete the population of stars having a
mass from $2-4\msun$. Therefore the difference between the observed
Pleiades mass function and the models in this mass range is more
likely due to binarity or a different IMF.

For model A the ratio is quite constant for less massive stars (and
BDs) thus demonstrating that the same fraction of objects has been
lost from the cluster independently of mass and thus {\it that energy
equipartition did not play a major role in de-populating the embedded
clusters but rather the violent process given by rapid gas expulsion}
during which the model clusters loose about 2/3 of their population.

The larger ratio for model~B comes about because this model is
initially more concentrated thus ultimately leading to a more massive
bound cluster because a larger fraction of the population evolves to a
more bound state due to energy equipartition before gas expulsion, as
already noted by KAH and considered in detail by Boily \& Kroupa
(2003a, 2003b). For model~B the ratio decreases from 0.45 at
$1\,M_\odot$ to 0.38 below $0.08\,\msun$. This comes about in this
model, which has a short equipartition time-scale, because
less-massive objects end up with a lower cluster binding energy due to
energy equipartition during the embedded phase. Gas expulsion then
leads to the low-mass objects being preferentially lost. The effect is
not very significant though, amounting to about 16~per cent.

%\begin{figure}[htbp]
%  \centering\leavevmode
%  \centerline{\includegraphics[width=\hsize]{ratio_sys0_100.ps}}
%  \caption{Fraction of systems remaining within the cluster after 100
%  Myr}
%  \label{sys_res}
%\end{figure}
%\begin{figure*}[htbp]
%  \centering\leavevmode
%  \includegraphics[width=0.45\hsize]{ratio_sys0_1.ps}
%  \includegraphics[width=0.45\hsize]{ratio_sys1_100.ps}
%  \caption{}
%  \label{sys_evol}
%\end{figure*}

\section{Conclusions}

The Pleiades is a young ($\approx 120$~Myr) close-by ($\approx
120$~pc) Galactic cluster and therefore allows very detailed
observational scrutiny. The available data together with theoretical
work suggest that the Pleiades may have been born similar to the ONC,
and that its subsequent evolution passed a violent phase as a result
of rapid gas expulsion once the massive cluster stars provided
sufficient feedback to the young nebula. The theoretical models
suggest that initially highly concentrated states (model~B), with
$\rho_{\rm C}\approx 10^{5.8}$~stars/pc$^3$, evolve to significant
mass segregation prior to gas expulsion, while less-concentrated
models with $\rho_{\rm C}\approx 10^{4.8}$~stars/pc$^3$ (model~A) do
not. As a result, the former looses slightly more of its low-mass
members relative to higher-mass ones than the latter
(Fig.~\ref{all_res}).

Both models are in good agreement with the observed MF in the Pleiades
(Fig.~\ref{mfAB}), although model~A yields a better fit which may
suggest that the Pleiades precursor had a concentration near
$\rho_{\rm C}\approx 10^{4.8}$~stars/pc$^3$ rather than being more
concentrated. The cluster's stellar IMF could thus have been very
similar to the Galactic-field average IMF for stars with $m\simless
2\,M_\odot$.  No significant stellar IMF variation can thus be
detected, in comparison to the Galactic-field. The sub-stellar part of
the Pleiades mass function also seems correctly reproduced by the
dynamical evolution of dense clusters. This suggests only marginal
evaporation of the lowest cluster members relative to higher mass
stars on a timescale of 100 Myr.

The Pleiades do appear to have a deficit of stars with $m\simgreat
2\,M_\odot$ (Fig.~\ref{mfAB}) compared to the number expected from the
average IMF indicating that the Pleiades IMF may have been steeper,
$\alpha_3 > 2.3$ for $m\ge2\,M_\odot$. This is interesting, because
there exists evidence that the stellar IMF corrected for binarity for
massive stars is probably steeper than a Salpeter IMF (Sagar \&
Richtler 1991; Kroupa \& Weidner 2003). These authors found $\alpha_3
\approx 2.7$ whereas an {\it apparent} Salpeter IMF indicates
$\alpha_{\rm app}=2.35$. However, it will be necessary to compute
additional models with different values for the star-formation
efficiency and mass to map-out feasible Pleiades precursors and the
range of physical conditions within the solution space. For example, a
lower initial density and thus a presumably larger loss of massive
stars during gas expulsion may reduce the discrepancy found here of
the measured MF and the standard IMF.

The Pleiades show very pronounced mass segregation
(Fig.~\ref{segreg100}) which is well reproduced by both models. This
indicates that the mass segregation that develops in model~B prior to
gas expulsion is forgotten by 100~Myr. The Pleiades is therefore in a
completely relaxed state, having almost completely lost memory of its
initial condition. These, however, are reflected in the properties of
its binary population, because the initial concentration limits the
binary-star periods that survive to old cluster age (KAH).

%%%%%%%%%%%%%%%%%%%%%%%%%%%%%%%%%%%%%%%%%%%%%%%%%%%%%%%%%%%%%%%%%%%%%

\begin{acknowledgements}

PK acknowledges support through DFG grant KR1635/4-1 and thanks the
staff of the Observatoire de Grenoble for their very kind hospitality
during the summer of~2002. This work made use of Aarseth's {\sc
Nbody6} code.

\end{acknowledgements}

%%%%%%%%%%%%%%%%%%%%%%%%%%%%%%%%%%%%%%%%%%%%%%%%%%%%%%%%%%%%%%%%%%%%%

\end{document}